\documentclass{article}
\usepackage{spconf,amsmath,graphicx,hyperref}
\usepackage[utf8]{inputenc}
\usepackage{soul}
\usepackage{stfloats}
\usepackage{placeins}

\usepackage{color}

\usepackage{color}

\title{When Noise Lowers the Loss: Rethinking Likelihood-Based \\Evaluation in Music Large Language Models}
\name{Xiaosha Li\textsuperscript{1}, \quad Chun Liu\textsuperscript{2}, \quad Ziyu Wang\textsuperscript{34}}
\address{\textsuperscript{1}Georgia Institute of Technology \
         \textsuperscript{2}ByteDance Inc.\\
         \textsuperscript{3}Courant Institute of Mathematical Sciences, New York University\\
         \textsuperscript{4} Mohamed bin Zayed University of Artificial Intelligence (MBZUAI)\\
         \texttt{xiaosha@gatech.edu},
         \texttt{chun.liu@bytedance.com}, \texttt{ziyu.wang@nyu.edu}
}
    
\begin{document}
 \ninept
\maketitle
\begin{abstract}
\label{sec:abstract}
The rise of music large language models (LLMs) demands robust methods of evaluating output quality, especially in distinguishing high-quality compositions from “garbage music”. Curiously, we observe that the standard cross-entropy loss---a core training metric---often decrease when models encounter systematically corrupted music, undermining its validity as a standalone quality indicator. To investigate this paradox, we introduce \textit{noise injection experiment}, where controlled noise signal of varying lengths are injected into musical contexts. 
We hypothesize that a model's loss reacting positively to these perturbations, specifically a sharp increase (``Peak'' area) for short injection, can serve as a proxy for its ability to discern musical integrity.
Experiments with MusicGen models in the audio waveform domain confirm that Music LLMs respond more strongly to local, texture-level disruptions than to global semantic corruption. 
Beyond exposing this bias, our results highlight a new principle: the shape of the loss curve---rather than its absolute value---encodes critical information about the quality of the generated content (i.e., model behavior). We envision this profile-based evaluation as a label-free, model-intrinsic framework for assessing musical quality—opening the door to more principled training objectives and sharper benchmarks.
\footnote{Code and demo page can be accessed via https://noiseloss.github.io.}

\end{abstract}

\section{Introduction}
\label{sec:intro}
In natural language processing, a common evaluation method is to feed text into a large language model (LLM) and compute its likelihood (or, equivalently, its loss) \cite{bengio2003neural}. The intuition is straightforward: A sequence assigned with a higher likelihood is considered more consistent with the model’s learned distribution, and therefore ``better''. With the rise of music LLMs, it is natural to extend this idea to music evaluation: input a musical sequence into the model and use likelihood as a proxy for quality.

\begin{figure}[htb]
\centering
\begin{minipage}[b]{1.0\linewidth}
  \centering
  \centerline{\includegraphics[width=8.5cm]{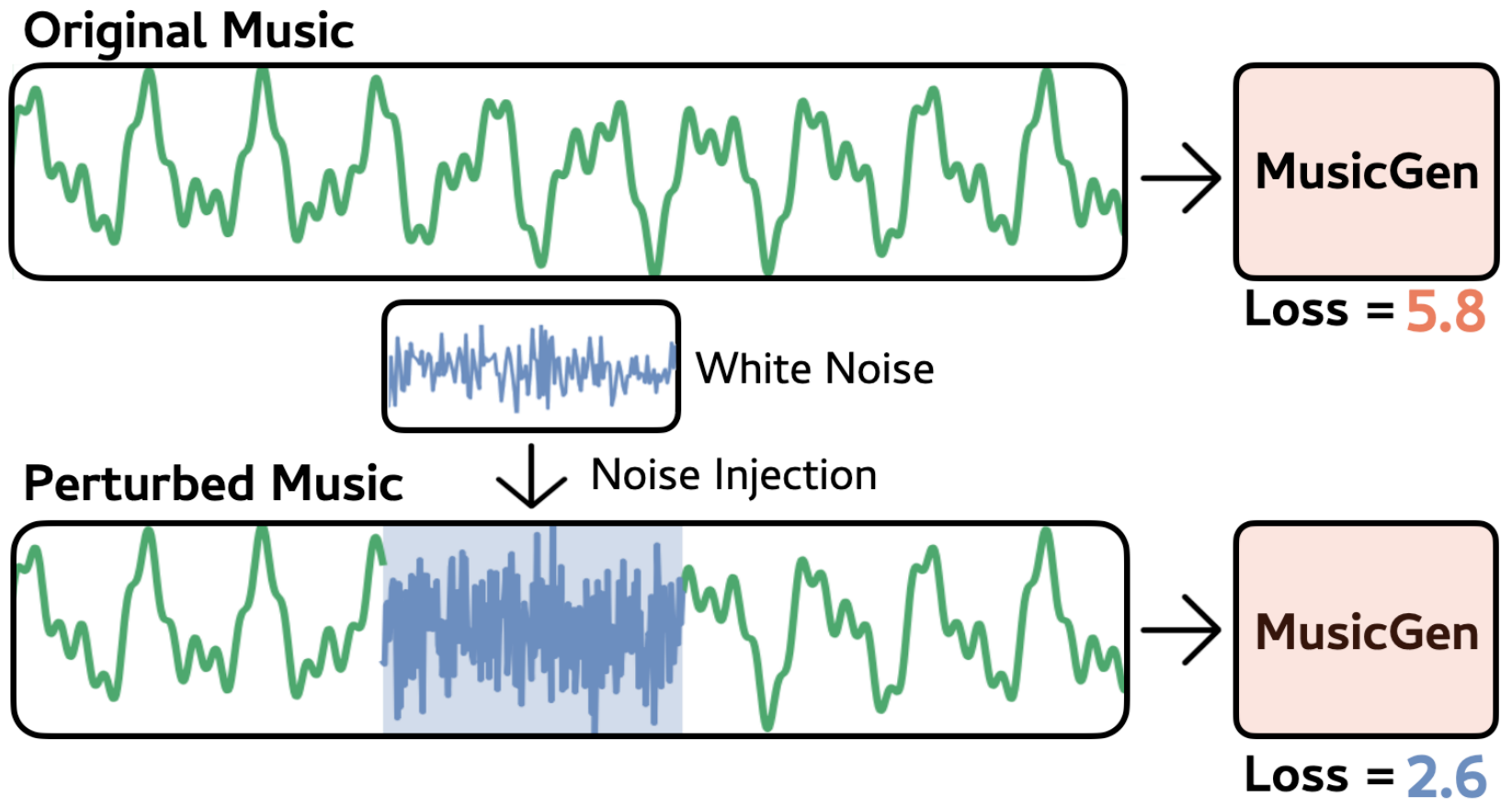}}
\end{minipage}
\caption{The figure shows that injecting noise into the input audio can unexpectedly reduce the loss; we refer to this phenomenon as the Context Amnesia Effect.}

\label{fig:figure1}
\end{figure}
However, the reliability of this approach in the music domain is far from established. While artifacts and biases of likelihood-based evaluation have been documented for text-based LLMs \cite{arora2022why,ranzato2016sequence,holtzman2020curious,ohi2024likelihood}, it remains unclear how these issues manifest in music or whether they align with human judgments of quality. To investigate this, we design a noise injection experiment in which perturbations (e.g. white noise) are added to musical sequences and the resulting changes in likelihood are measured (Figure~\ref{fig:figure1}). Intuitively, one would expect that corrupting music with noise decreases its likelihood, thereby increasing the loss. Surprisingly, we observe the opposite: adding noise frequently \textit{lowers} the loss.

Our analysis reveals that the unexpected reduction in loss under noise arises from changes in per-token likelihood. At the onset of a noise segment, the model reacts with a sharp spike in loss, signaling recognition of inconsistency with the preceding context. Yet almost immediately afterward, the loss drops and remains low for the duration of the noise, regardless of the original context. This happens because music LLMs already assign relatively low loss to certain forms of noise, whose regularity makes it easier to predict than real music. Once the perturbation ends, the loss realigns with the original context, but with significantly higher variance. In other words, the model briefly “resists” the disturbance, then readily “forgets” the prior musical material and adapts to the noise as if it were the prevailing context. This behavior emerges consistently across noise types, musical styles, and transformer-based model variants. We refer to this phenomenon as the \textit{Context Amnesia Effect}.

\begin{figure*}[t]
  \centering
  \includegraphics[width=\linewidth]{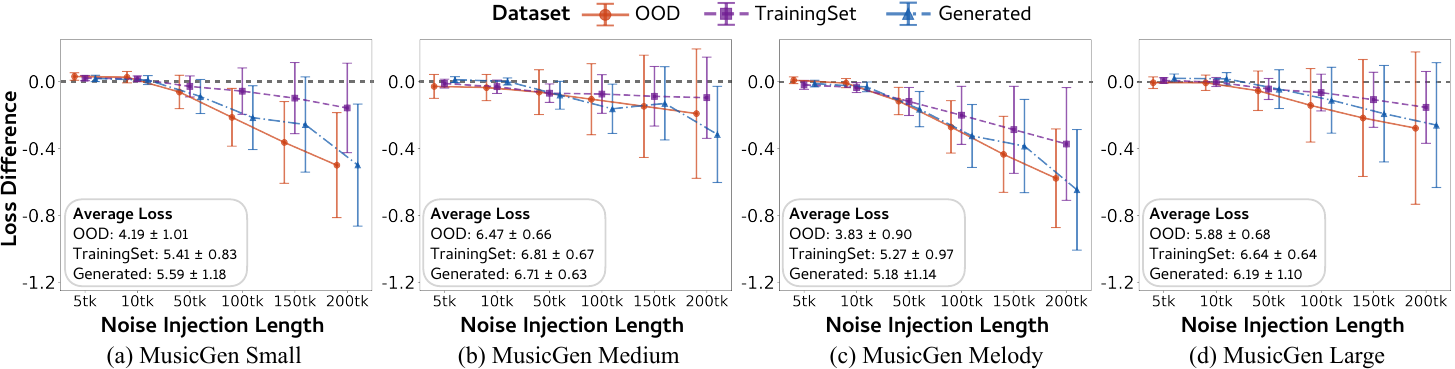}
  \caption{Comparison of model performance under white-noise injection on the dataset, with standard deviation measured across songs.}
  \label{fig:figure2}
\end{figure*}

This finding highlights a fundamental limitation of likelihood-based evaluation in music. When measured by loss, LLMs can reliably detect only very short-term inconsistencies (e.g., onset noise) but fail to register longer-term structural degradations (e.g., phrase reorderings). In fact, the loss response to perturbations is highly inconsistent: it may rise, fall, or remain unchanged, making it an unreliable indicator of musical quality. This unpredictability is a concrete manifestation of \emph{exposure bias}\cite{ranzato2016sequence} in music LLMs.

At the same time, our results point to a more promising direction. While absolute loss values are uninformative, the local dynamics of the loss curve carry meaningful signals. In particular, we consistently observe a sharp peak at the onset of perturbation, followed by assimilation and recovery phases, which appear more reliable than raw likelihood in capturing model behavior. This suggests that profile-based evaluation focusing on the shape of token-wise loss curves rather than absolute values may provide a stronger foundation for future methods of automatic music evaluation.


\section{Related Work}
\label{sec:related work}
LLM-based evaluation leverages large models to assess other models' outputs through several major paradigms. Automatic methods include likelihood-based metrics, which measure probability fit but often neglect content quality \cite{ohi2024likelihood}, and prompt-based multiple-choice tests that expose paradoxes in evaluative ability \cite{west2023paradox}. In contrast, non-automatic approaches such as human preference arenas rank models via large-scale pairwise voting \cite{chiang2024chatbotarena}. Recent extensions like G-Eval and GPTScore, employ LLMs directly as judges—either through structured reasoning with probability-weighted scoring \cite{liu2023geval} or flexible instruction-based evaluation \cite{fu2023gptscore}—and report stronger correlations with human ratings in language tasks. However, applying these methods to music remains challenging: musical semantics are ambiguous, salient moments and long-range structure are hard to capture, and high-probability continuations often fail to reflect quality.

Automatic evaluation offers scalability and reduced cost, but its reliability remains uncertain. Although likelihood is an effective training objective, it suffers from exposure bias and objective mismatch at inference time \cite{ranzato2016sequence,xu2024crossentropy}. Using likelihood for decoding further yields bland and repetitive text, diverging from the creativity and diversity valued in human language \cite{holtzman2020curious}. LLM-based evaluators also exhibit likelihood bias, overrating high-likelihood but superficial outputs \cite{ohi2024likelihood}, and suffer from inconsistency, with familiarity and anchoring effects and sensitivity to prompts that do not affect human judgment \cite{stureborg2025biased}. More broadly, such artifacts echo shortcut learning, where models exploit spurious correlations rather than genuine ability, as observed across vision and language domains \cite{geirhos2020shortcut,srivastava2024shortcuts}.

In music, evaluation continues to rely primarily on human ratings \cite{groetschla2025benchmarking}, with objective metrics emerging only recently \cite{lerch2025survey}. Fr{\'e}chet Audio Distance (FAD) and MAD probe fidelity, musicality, and diversity \cite{gui2024fad,huang2024mad}, while platforms like Music Arena scale evaluation via listener preferences \cite{kim2025musicarena}. CMI-Bench reframes music understanding into instruction-following tasks \cite{wang2025cmibench}. Audiobox-Aesthetics represents an early attempt to integrate human ratings into automated evaluation through a weighted model, though its dimensions remain limited \cite{tjandra2025audiobox}. Unlike in the language domain, no established methodology exists for using large language models as automatic judges of music, motivating our investigation.

\section{Noise Injection Experiment}
\label{sec:experiment}
In this section, we introduce the noise injection experiment, which demonstrates the counterintuitive effect that model prediction loss decreases (or equivalently, likelihood increases) when perturbations are applied to the input audio. The experimental setup is described in Section~\ref{sec:3:setting}, and the results are presented in Section~\ref{sec:3:result}.

\subsection{Experiment Setting}
\label{sec:3:setting}
Given an audio signal $x$ tokenized as $x_{1:T}$, where $T$ is the sequence length (e.g., via EnCodec~\cite{defossez2023encodec} in MusicGen), a generative model computes the loss $\ell(x_{1:T})$ as the negative log-likelihood:

\begin{equation}
\ell(x_{1:T}) = - \sum_{t=1}^{T} \log p_\theta(x_t \mid x_{<t}),
\end{equation}
We define the perturbed signal $x'_{1: T}$ as 
\begin{equation}
x'_i = 
\begin{cases}
x_i, & i \notin \mathcal{I}, \\
\epsilon_i, & i \in \mathcal{I},
\end{cases}
\end{equation}
where $\mathcal{I}$ is the perturbed time steps and $\epsilon_i$ denotes the given noise. Similarly, the perturbed sequence $x'_{1:T}$ has the loss
\begin{equation}
\ell(x'_{1:T}) = - \sum_{t=1}^{T} \log p_\theta(x'_t \mid x'_{<t}).
\end{equation}
We also define the \textit{token-wise loss difference} at time step $t$ as
\begin{equation}
\Delta \ell_t = -\log p_{\theta}(x'_t|x'_{<t}) + \log p_{\theta}(x_t|x_{<t}),
\label{eq:token-wise-diff}
\end{equation}
which measures the change of sequence likelihood under perturbation from that of the original sequence at token $t$.
We first use white noise with controlled loudness ($-30$ to $-12$ dB) as injected noise, where the noise loudness is matched to the original audio to account for variations in overall amplitude across datasets, as we empirically observe that the resulting loss magnitude is sensitive to the noise loudness level.
For an audio consisting of 750 tokens (15 seconds), the noise is injected at the 250-th token (5 seconds), which provides sufficient history for the model to establish context.
Here, tokens correspond to RVQ-based audio tokens at a rate of 50~Hz, meaning that each token represents approximately 20~ms of audio (with all signals resampled to 32~kHz).
Perturbation lengths are set to $5,\allowbreak 10,\allowbreak 50,\allowbreak 100,\allowbreak 150,\allowbreak 200$ tokens, corresponding to $0.1,\allowbreak 0.2,\allowbreak 1.0,\allowbreak 2.0,\allowbreak 3.0,\allowbreak 4.0$ seconds.
The chosen lengths cover multiple levels of musical perturbation, disrupting semantics at approximately the frame, note, beat, and measure levels.

\subsection{Experiment Results}
\label{sec:3:result}
In our experiments, we evaluate noise injection on three types of music data: (1) \textbf{TrainingSet}: a subset of the ShutterStock~\footnote{https://www.shutterstock.com} training corpus used for MusicGen, consisting of 20 songs; (2) \textbf{Generated}: 140 samples produced by MusicGen-Small under broad generation settings (top-k = 10, 50, 100, 150, 200, 250, 500, etc); and (3) \textbf{Out of Distribution (OOD)}: 78 classical pieces from the ASAP dataset~\footnote{https://github.com/fosfrancesco/asap-dataset}, spanning a wide range of composers and styles. We evaluate autoregressive LLMs in waveform, including MusicGen (Small(300M)\allowbreak / Medium(1.5B)\allowbreak / Large(3.3B)\allowbreak / Melody(1.5B)) models \cite{copet2023simple}.  

We compute the loss difference $\Delta\ell = \ell(x'_{1:T}) - \ell(x_{1:T})$, and the results are presented in Fig.~\ref{fig:figure2}. Across all models, datasets, and noise lengths, we observe a consistent pattern: when the injected noise is short, the loss difference remains close to zero; as the noise length increases, the loss difference becomes negative, indicating that longer perturbations systematically decrease the loss.

To verify the robustness of this trend, we apply both Pearson and Spearman correlation tests between perturbation length and average loss difference.
In the large-sample setting, where each data point represents a specific $(\text{model}, \text{dataset}, \text{noise-length})$ combination ($\approx 78$ points total), both Pearson and Spearman correlation coefficients are strongly negative ($r < -0.85$, $p < 0.001$), confirming a highly significant negative trend.
Even in the small-sample setting, with results aggregated by noise length ($\approx 6$ groups), correlations remain robust ($r < -0.91$) and statistically significant ($p < 0.05$), a finding further corroborated by the significantly negative slopes observed in linear regression tests.

To further examine the generalizability of this effect, we repeat the analysis on the YuE (1B) \cite{yue2025scaling} autoregressive model using three datasets: TrainingSet, OOD, and Generated data produced by YuE itself. For both TrainingSet and OOD data, Pearson correlation tests reveal a consistent and significant negative relationship between noise length and loss difference ($r < -0.90$, $p < 0.01$), corroborated by significantly negative regression slopes. For the Generated data, although no statistically significant correlation between noise length and loss difference is observed ($p > 0.05$), all noise-injected samples consistently exhibit lower loss values compared to their corresponding clean generated counterparts. This generalized finding indicates that the absolute loss value alone is insufficient to determine whether a musical sequence has been perturbed by noise, particularly in model-generated music. Visualizations of these experiments are available on our demo page.


\section{Analysis of Loss Dynamics}
\label{sec:analysis}
We now analyze in detail how loss behaves in the noise injection experiment. Our goals are twofold: first, to explain the counterintuitive trends observed in Section~\ref{sec:experiment} and second, to quantify how perturbations reshape token-wise loss dynamics. We first use per-token visualization to highlight a consistent three-stage effect (Section~\ref{subs: Analysis of token-wise loss}), then validate this effect with an automatic region-detection experiment (Section~\ref{subs:validation}).

\subsection{Token-Wise Loss Dynamics}
\label{subs: Analysis of token-wise loss}
\begin{figure}[t]
\centering
\begin{minipage}[b]{1.0\linewidth}
  \centering
  \centerline{\includegraphics[width=8.5cm]{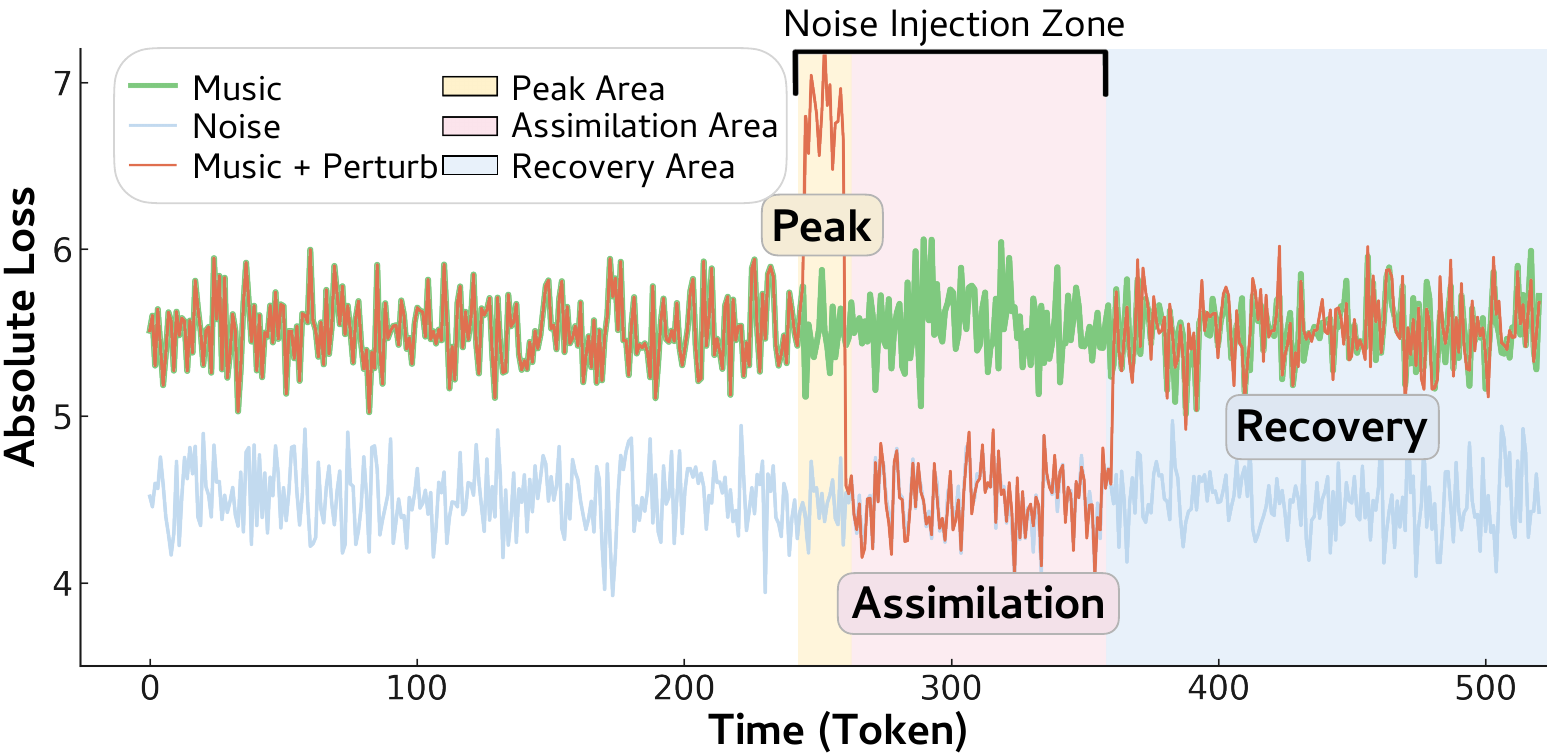}}
\end{minipage}
\caption{Loss curve of noise injection experiment.}
\label{fig:figure3}
\end{figure}

To investigate the counterintuitive behavior, we analyze the noise injection setting in Eq.~\eqref{eq:token-wise-diff} at the per-token level. We compute the loss difference $\Delta \ell_t$ and visualize token-wise loss curves for original music, pure noise, and music+perturb across all models and datasets in our experiments (Fig.~\ref{fig:figure3}), revealing how perturbation reshapes the loss dynamics. 

\FloatBarrier
\begin{figure}[htb]
\centering
\begin{minipage}[b]{1.0\linewidth}
  \centering
  \centerline{\includegraphics[width=8.5cm]{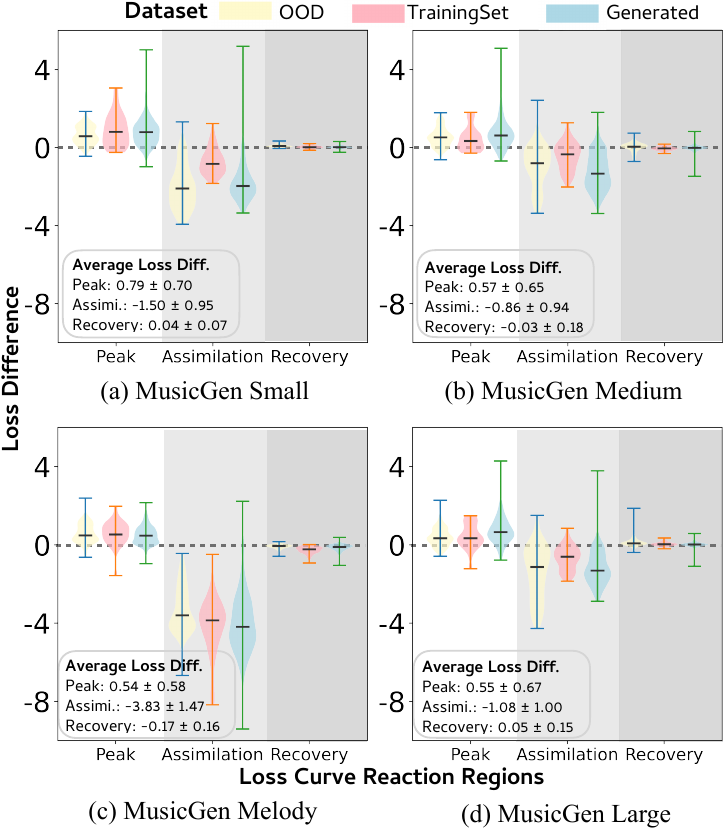}}
\end{minipage}
\caption{Three-stage behavior after noise injection.}
\label{fig:figure4}
\end{figure}

Across different perturbation lengths, we observe three characteristic regions in the token-wise loss trajectory:

\begin{itemize}
    \item \textbf{Peak area:} In the first $\sim$100 ms ($\approx$5 tokens), loss spikes due to local inconsistency with the preceding context.
    \item \textbf{Assimilation area:} Within the perturbation window, loss rapidly decreases and stabilizes at a low value, largely insensitive to context.
    \item \textbf{Recovery area:} After the perturbation ends, predictions become unstable and loss oscillates around the baseline, reflecting exposure bias.
\end{itemize}

This three-stage behavior suggests that the model transiently loses effective access to prior musical context, as if earlier information were forgotten, which we term the Context Amnesia Effect.

\subsection{Validation Through Automated Region Detection}
\label{subs:validation}
To confirm that these three regions are consistent, we conduct a region-detection experiment. We apply a moving average to $\Delta \ell_t$ to suppress local fluctuations and automatically identify boundaries where the loss crosses zero for five consecutive tokens. This procedure robustly detects the onset peak, the assimilation plateau, and the unstable recovery phase (Fig.~\ref{fig:figure4}).  

These results validate the hypothesis suggested by the qualitative analysis in Section~\ref{subs: Analysis of token-wise loss}: models respond strongly to short, local perturbations but fail to reflect long-range structural disruptions. This Context Amnesia Effect underlies the unreliability of loss-based evaluation for music.

\section{Discussion}
\label{sec:discussion}
In the previous sections, we demonstrated that adding white noise counterintuitively lowers the loss and introduced the Context Amnesia Effect to explain this behavior. While these results clarify why model loss fails to capture long-range structural disruptions, it remains unclear whether this limitation is unique to noise injection or reflects a more general weakness of loss as an evaluation metric. In this section, we extend our analysis to more realistic settings specifically, those where model loss would naturally be used to assess musical quality. 
We first examine order shuffling as an alternative perturbation in Section~\ref{sebsuc1:shuffle}, showing that the same patterns of short-range sensitivity and long-range insensitivity persist. We then relate this phenomenon to the broader concept of exposure bias in Section~\ref{sebsuc2: Relation to Exposure Bias}, highlighting its implications for evaluating generative music models.

\subsection{Alternative Perturbation: Order Shuffling}
\label{sebsuc1:shuffle}

To further examine our findings, we extend the analysis to another perturbation: order shuffling. Unlike noise injection, shuffling preserves the same amount of information while disrupting musical order in a structured manner. We shuffle segments of different lengths (1, 2, 5, 10, 35, 50, 70, 100, 150, and 200 tokens) to cover a broad range of perturbation scales, from local rearrangements to long-range disruptions of musical structure.

\begin{figure}[t]
  \centering
  \centerline{\includegraphics[width=8.5cm]{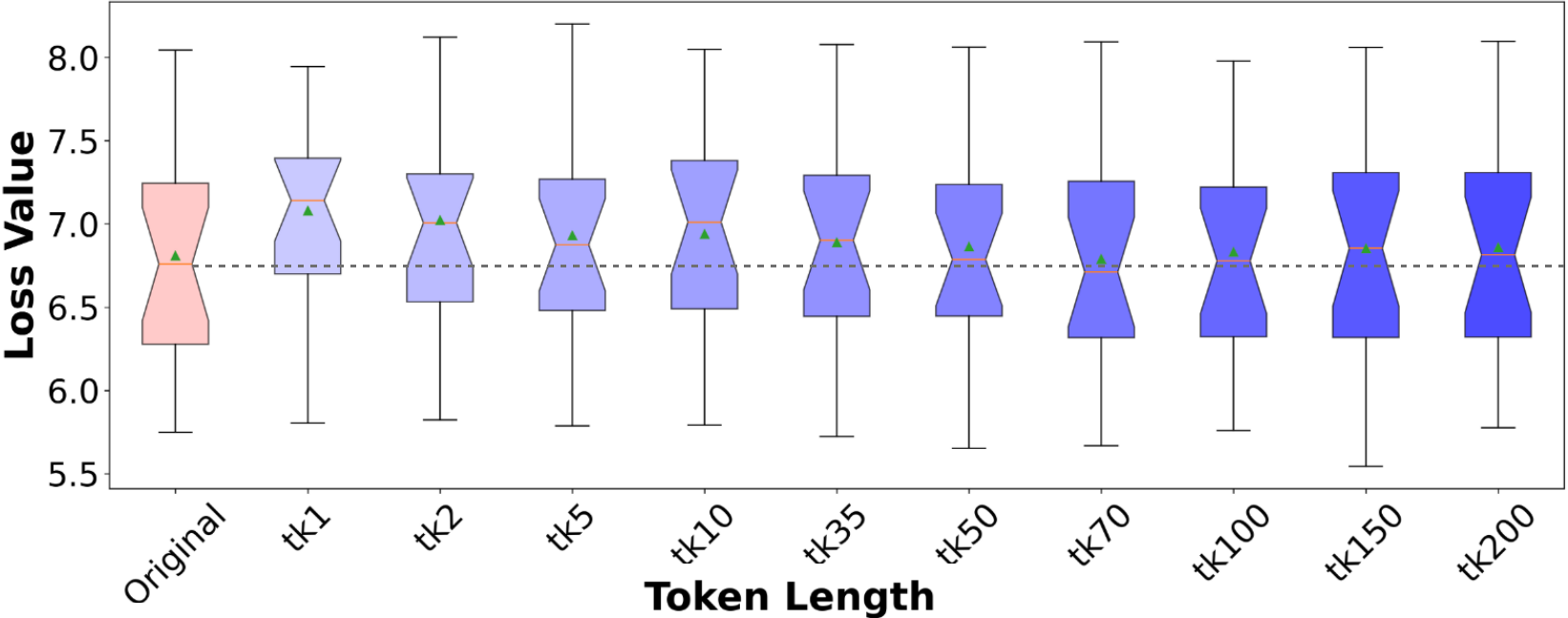}}
\caption{Example of shuffle order perturbation with experimental results.}
\label{fig:figure5}
\end{figure}

As shown in Fig.~\ref{fig:figure5}, the results closely mirror those of noise injection. Short shuffles produce a sharp spike in loss, as the perturbation is dominated by the peak region. As the shuffle length increases, the assimilation region becomes longer, during which the model adapts to the perturbed order and effectively loses access to the original context, causing the loss to remain close to the baseline. Although long-span shuffling introduces substantial and structured degradation of musical form, the global loss remains largely unchanged, indicating the unreliability of global loss for evaluating model performance under long-range structural disruptions.

\subsection{Relation to Exposure Bias}
\label{sebsuc2: Relation to Exposure Bias}
The Context Amnesia Effect is closely related to the well-known phenomenon of exposure bias, where auto-regressive models struggle to recover after errors during inference-time generation. In our setting, injected noise produces tokens that are out-of-distribution relative to the training data, leading to a brief loss spike at noise onset followed by rapid adaptation and increasingly unstable predictions. After such errors, the model effectively shortens its usable context and relies on corrupted input, indicating that exposure bias not only degrades generation but also undermines the reliability of likelihood-based evaluation. In music, where surprise, tension, and novelty are essential, unfamiliar yet meaningful passages may therefore be misinterpreted as perturbations and undervalued in the same manner as noise.

Our findings suggest current LLMs cannot reliably use absolute loss to distinguish between works of differing quality (e.g., between canonical classical compositions and generic training-set samples). We encourage future research to further explore the depth of this limitation in music evaluation.

\section{Conclusion}
\label{sec:conclusion}
In this work, we investigated the reliability of loss-based evaluation for music LLMs, anchored by a counterintuitive phenomenon observed in our noise injection experiment. We identified a key pattern termed the Context Amnesia Effect: when perturbations are introduced, the loss curve consistently exhibits three characteristic regions—\textbf{Peak}, \textbf{Assimilation}, and \textbf{Recovery}. By visualizing loss curves and token-wise loss differences, we demonstrated how models detect only instantaneous inconsistencies while failing to respond to longer-term structural changes. Extending our analysis to additional perturbations, such as order shuffling, further confirmed that absolute loss is unreliable for evaluating musical quality, especially at the compositional level.  


Our findings underscore a fundamental limitation of likelihood-based evaluation for music: absolute loss values cannot function as a stable indicator of musical quality. Instead, the shape and local dynamics of the loss curve—particularly the onset peak—offer clearer, more consistent signals. We frame this profile-based perspective as an initial step toward developing more reliable automatic evaluation frameworks, ones that better align with music’s unique structural characteristics.

\section{Acknowledgment}
The authors would like to thank Prof. Alexander Lerch and Yikang Shen for their valuable guidance, and Wenye Ma for her helpful assistance.


\vfill\pagebreak



\bibliographystyle{IEEEbib}
\bibliography{strings,refs}

@article{bengio2003neural,
  title={A {Neural} {Probabilistic} {Language} {Model}},
  author={Bengio, Yoshua and Ducharme, Réjean and Vincent, Pascal and Jauvin, Christian},
  journal={Journal of Machine Learning Research},
  year={2003}
}

@article{arora2022why,
  title={Why {Exposure} {Bias} {Matters}: An {Imitation} {Learning} {Perspective} of {Error} {Accumulation} in {Language} {Generation}},
  author={Arora, Kushal and El Asri, Layla and Bahuleyan, Hareesh and Cheung, Jackie Chi Kit},
  journal={Findings of the Association for Computational Linguistics (ACL)},
  year={2022},
  doi={10.48550/arXiv.2204.01171},
  eprint={2204.01171},
  archivePrefix={arXiv},
  primaryClass={cs.CL},
  url={https://arxiv.org/abs/2204.01171}
}

@inproceedings{ranzato2016sequence,
  title={Sequence {Level} {Training} with {Recurrent} {Neural} {Networks}},
  author={Ranzato, Marc’Aurelio and Chopra, Sumit and Auli, Michael and Zaremba, Wojciech},
  booktitle={International Conference on Learning Representations (ICLR)},
  year={2016},
  url={https://arxiv.org/abs/1511.06732}
}

@inproceedings{holtzman2020curious,
  title={The {Curious} {Case} of {Neural} {Text} {Degeneration}},
  author={Holtzman, Ari and Buys, Jan and Du, Li and Forbes, Maxwell and Choi, Yejin},
  booktitle={International Conference on Learning Representations (ICLR)},
  year={2020}
}

@inproceedings{ohi2024likelihood,
  title={Likelihood-based {Mitigation} of {Evaluation} {Bias} in {Large} {Language} {Models}},
  author={Ohi, Masanari and Kaneko, Masahiro and Koike, Ryuto and Loem, Mengsay and Okazaki, Naoaki},
  booktitle={Findings of the Association for Computational Linguistics (ACL)},
  pages={3237--3245},
  year={2024},
  url={https://aclanthology.org/2024.findings-acl.191}
}

@inproceedings{xu2024crossentropy,
  title={Understanding the {Role} of {Cross-Entropy} {Loss} in {Fairly}
{Evaluating} {Large} {Language} {Model-based} {Recommendation}},
  author={Xu, Cong and Zhu, Zhangchi and Wang, Jun and Wang, Jianyong and Zhang, Wei},
  booktitle={Proceedings of the ACM Conference on Recommender Systems (RecSys)},
  year={2024},
  url={https://arxiv.org/pdf/2402.06216}
}

@inproceedings{west2023paradox,
  title={{The {Generative} {AI} {Paradox}}: {What} {It} {Can} {Create}, {It} {May} {Not} {Understand}},
  author={West, Peter and Lu, Ximing and Dziri, Nouha and Brahman, Faeze and Li, Linjie and Hwang, Jena D. and Jiang, Liwei and Fisher, Jillian and Ravichander, Abhilasha and Chandu, Khyathi Raghavi and Newman, Benjamin and Koh, Pang Wei and Ettinger, Allyson and Choi, Yejin},
  booktitle={International Conference on Learning Representations (ICLR)},
  year={2024},
  url={https://arxiv.org/abs/2311.00059}
}

@inproceedings{chiang2024chatbotarena,
  title={Chatbot {Arena}: {An} {Open} {Platform} for {Evaluating} {LLMs} by {Human} {Preference}},
  author={Chiang, Wei-Lin and Zheng, Lianmin and Zhuang, Siyuan and Wallace, Eric and Li, Tianjun and Sheng, Yingbo and Wu, Rose and others},
  booktitle={Proceedings of the 2024 Conference on Empirical Methods in Natural Language Processing (EMNLP)},
  year={2024},
  url={https://arxiv.org/abs/2403.04132}
}

@inproceedings{liu2023geval,
  title={{G-Eval}: {NLG} {Evaluation} using {GPT-4} with {Better} {Human} {Alignment}},
  author={Liu, Yang and Iter, Dan and Xu, Yichong and Wang, Shuohang and Xu, Ruochen and Zhu, Chenguang},
  booktitle={Proceedings of the 2023 Conference on Empirical Methods in Natural Language Processing (EMNLP)},
  pages={2511--2522},
  year={2023},
  url={https://aclanthology.org/2023.emnlp-main.153}
}

@inproceedings{fu2023gptscore,
  title={{GPTScore}: {Evaluate} as {You} {Desire}},
  author={Fu, Jinlan and Ng, See-Kiong and Jiang, Zhengbao and Liu, Pengfei},
  booktitle={Proceedings of the 2024 Conference of the North American Chapter of the Association for Computational Linguistics: Human Language Technologies (NAACL)},
  pages={6556--6576},
  year={2024},
  url={https://aclanthology.org/2024.naacl-long.365}
}

@article{stureborg2025biased,
  title={{Large} {Language} {Models} are {Inconsistent} and {Biased} {Evaluators}},
  author={Stureborg, Rickard and Alikaniotis, Dimitris and Suhara, Yoshi},
  journal={arXiv preprint arXiv:2405.01724},
  year={2024},
  url={https://arxiv.org/pdf/2405.01724}
}

@article{geirhos2020shortcut,
  title={Shortcut {Learning} in {Deep} {Neural} {Networks}},
  author={Geirhos, Robert and Jacobsen, J{\"o}rn-Henrik and Michaelis, Claudio and Zemel, Richard and Brendel, Wieland and Bethge, Matthias and Wichmann, Felix A.},
  journal={Nature Machine Intelligence},
  volume={2},
  number={11},
  pages={665--673},
  year={2020}
}

@inproceedings{srivastava2024shortcuts,
  title={Do {LLMs} {Overcome} {Shortcut} {Learning}? {An} {Evaluation} of {Shortcut} {Challenges} in {Large} {Language} {Models}},
  author={Yuan, Yu and Zhao, Lili and Zhang, Kai and Zheng, Guangting and Liu, Qi},
  booktitle={Proceedings of the 2024 Conference on Empirical Methods in Natural Language Processing (EMNLP)},
  pages={12188--12200},
  year={2024}
}

@inproceedings{groetschla2025benchmarking,
  title={{Benchmarking Music Generation Models and Metrics via Human Preference Studies}},
  author={Gr{\"o}tschla, Florian and Solak, Ahmet and Lanzend{\"o}rfer, Luca A. and Wattenhofer, Roger},
  booktitle={2025 IEEE International Conference on Acoustics, Speech and Signal Processing (ICASSP)},
  year={2025},
  publisher={IEEE},
  doi={10.1109/ICASSP49660.2025.10887745}
}

@article{lerch2025survey,
  title={Survey on the {Evaluation} of {Generative} {Models} in {Music}},
  author={Lerch, Alexander and Arthur, Claire and Bryan-Kinns, Nick and Ford, Corey and Sun, Qianyi and Vinay, Ashvala},
  journal={ACM Computing Surveys},
  volume={58},
  number={4},
  pages={99},
  year={2025},
  url={https://dl.acm.org/doi/10.1145/3769106}
}

@inproceedings{gui2024fad,
  title={Adapting {Fr{\'e}chet} {Audio} {Distance} for {Generative} {Music} {Evaluation}},
  author={Gui, Azalea and Gamper, Hannes and Braun, Sebastian and Emmanouilidou, Dimitra},
  booktitle={2024 IEEE International Conference on Acoustics, Speech and Signal Processing (ICASSP)},
  pages={1331--1335},
  year={2024},
  publisher={IEEE}
}

@inproceedings{huang2024mad,
  title={Aligning {Text-to-Music} {Evaluation} with {Human} {Preferences}},
  author={Huang, Yichen and Donahue, Chris},
  booktitle={International Society for Music Information Retrieval Conference (ISMIR)},
  year={2025},
  url={https://arxiv.org/abs/2503.16669}
}

@inproceedings{kim2025musicarena,
  title={Music {Arena}: {Live} {Evaluation} for {Text-to-Music}},
  author={Kim, Yonghyun and Chi, Wayne and Angelopoulos, Anastasios N. and Chiang, Wei-Lin and Saito, Koichi and Watanabe, Shinji and Mitsufuji, Yuki and Donahue, Chris},
  booktitle={Advances in Neural Information Processing Systems (NeurIPS)},
  year={2025},
  url={https://arxiv.org/abs/2507.20900}
}

@inproceedings{wang2025cmibench,
  title={{CMI-Bench}: {A Comprehensive Benchmark for Evaluating Music Instruction Following}},
  author={Wang, Yuxuan and Sun, Ming and Chen, Hao and Zhang, Rui and Agres, Kat and Yang, Yi-Hsuan},
  booktitle={International Society for Music Information Retrieval Conference (ISMIR)},
  year={2025},
  url={https://arxiv.org/abs/2506.12285}
}

@article{tjandra2025audiobox,
  title={{Meta Audiobox Aesthetics}: {Unified Automatic Quality Assessment for Speech, Music, and Sound}},
  author={Tjandra, Andros and Wu, Yi-Chiao and Guo, Baishan and Hoffman, John and Ellis, Brian and Vyas, Apoorv and Shi, Bowen and Chen, Sanyuan and Le, Matt and Zacharov, Nick and Wood, Carleigh and Lee, Ann and Hsu, Wei-Ning},
  journal={arXiv preprint arXiv:2502.05139},
  year={2025},
  url={https://arxiv.org/pdf/2502.05139}
}

@article{defossez2023encodec,
  title={High Fidelity Neural Audio Compression},
  author={D{\'e}fossez, Alexandre and Copet, Jade and Synnaeve, Gabriel and Adi, Yossi},
  journal={Transactions on Machine Learning Research (TMLR)},
  year={2023},
  note={Accepted September 2023},
  url={https://openreview.net/forum?id=ivCd8z8zR2}
}

@inproceedings{copet2023simple,
  title={{Simple and Controllable Music Generation}},
  author={Copet, Jade and Kreuk, Felix and Gat, Itai and Remez, Tal and Kant, David and Synnaeve, Gabriel and Adi, Yossi and D{\'e}fossez, Alexandre},
  booktitle={Proceedings of the 37th Conference on Neural Information Processing Systems (NeurIPS)},
  year={2023}
}

@article{yue2025scaling,
  title   = {YuE: {Scaling} {Open} {Foundation} {Models} for {Long-Form} {Music} {Generation}},
  author  = {Yuan, Ruibin and Lin, Hanfeng and Guo, Shuyue and Zhang, Ge and Pan, Jiahao and Zang, Yongyi and Liu, Haohe and Liang, Yiming and Ma, Wenye and Du, Xingjian and others},
  journal = {arXiv preprint arXiv:2503.08638},
  year    = {2025},
  url     = {https://arxiv.org/abs/2503.08638},
}

\end{document}